\documentclass[aps,prl,twocolumn,10pt,amsmath,superscriptaddress,amssymb,longbibliography]{revtex4-1}

\usepackage{booktabs}
\usepackage{multirow}

\usepackage{epsfig}
\usepackage{threeparttable}
\usepackage{epstopdf}
\usepackage{dcolumn}
\usepackage{amsbsy}
\usepackage{bm}
\usepackage{color,CJK}
\usepackage{amsmath,amssymb,amsfonts}
\usepackage{appendix}
\usepackage{graphicx}
\usepackage{algorithm}
\usepackage{enumerate}
\usepackage{makeidx}
\usepackage{braket}
\usepackage[usenames,dvipsnames]{xcolor}
\usepackage[driverfallback=dvipdfmx,colorlinks,linkcolor=blue,citecolor=blue,urlcolor=blue]{hyperref}

\begin{document}

\title{Experimental study of tune-out wavelengths for spin-dependent optical lattice in $^{87}$Rb Bose-Einstein condensation}

\author{Kai Wen}
\affiliation{State Key Laboratory of Quantum Optics and Quantum
Optics Devices, \\ Institute of Opto-electronics, Shanxi University,
Taiyuan, Shanxi 030006, People's Republic of China }
\affiliation{Collaborative Innovation Center of Extreme Optics,
Shanxi University, Taiyuan, Shanxi 030006, People's Republic of
China }

\author{Zengming Meng}
\email[Corresponding author email: ]{zmmeng01@sxu.edu.cn; }
\affiliation{State Key Laboratory of Quantum Optics and Quantum
Optics Devices, \\ Institute of Opto-electronics, Shanxi University,
Taiyuan, Shanxi 030006, People's Republic of China }
\affiliation{Collaborative Innovation Center of Extreme Optics,
Shanxi University, Taiyuan, Shanxi 030006, People's Republic of
China }

\author{Liangwei Wang}
\affiliation{State Key Laboratory of Quantum Optics and Quantum
Optics Devices, \\ Institute of Opto-electronics, Shanxi University,
Taiyuan, Shanxi 030006, People's Republic of China }
\affiliation{Collaborative Innovation Center of Extreme Optics,
Shanxi University, Taiyuan, Shanxi 030006, People's Republic of
China }

\author{Liangchao Chen}
\affiliation{State Key Laboratory of Quantum Optics and Quantum
Optics Devices, \\ Institute of Opto-electronics, Shanxi University,
Taiyuan, Shanxi 030006, People's Republic of China }
\affiliation{Collaborative Innovation Center of Extreme Optics,
Shanxi University, Taiyuan, Shanxi 030006, People's Republic of
China }

\author{Lianghui Huang}
\affiliation{State Key Laboratory of Quantum Optics and Quantum
Optics Devices, \\  Institute of Opto-electronics, Shanxi
University, Taiyuan, Shanxi 030006, People's Republic of China }
\affiliation{Collaborative Innovation Center of Extreme Optics,
Shanxi University, Taiyuan, Shanxi 030006, People's Republic of
China }

\author{Pengjun Wang}
\affiliation{State Key Laboratory of Quantum Optics and Quantum
Optics Devices, \\ Institute of Opto-electronics, Shanxi University,
Taiyuan, Shanxi 030006, People's Republic of China }
\affiliation{Collaborative Innovation Center of Extreme Optics,
Shanxi University, Taiyuan, Shanxi 030006, People's Republic of
China }

\author{Jing Zhang}
\email[Corresponding author email: ]{jzhang74@yahoo.com;
\\jzhang74@sxu.edu.cn}
\affiliation{State Key Laboratory of Quantum Optics and Quantum
Optics Devices, \\ Institute of Opto-electronics, Shanxi University,
Taiyuan, Shanxi 030006, People's Republic of China }
\affiliation{Collaborative Innovation Center of Extreme Optics,
Shanxi University, Taiyuan, Shanxi 030006, People's Republic of
China }


\begin{abstract}
We study the periodic potential of one-dimensional optical lattice originated from scalar shift and vector shift by manipulating the lattice polarizations. The ac Stark shift of optical lattice is measured by Kapitza-Dirac scattering of $^{87}$Rb Bose-Einstein condensate and the characteristics of spin-dependent optical lattice are presented by scanning the lattice wavelength between the D1 and D2 lines. At the same time, tune-out wavelengths that ac Stark shift cancels can be probed by optical lattice. We give the tune-out wavelengths in more general cases of balancing the contributions of both the scalar and vector shift. Our results provide a clear interpretation for spin-dependent optical lattice and tune-out wavelengths, and help to design it by choosing the appropriate lattice wavelength.
\end{abstract}

\maketitle
\section{Introduction}

Optical lattice for ultracold atoms has become an increasingly important technology in many-body physics~\cite{RMP1}, quantum simulation, quantum computation, quantum information storage and high precision measurements~\cite{aidelsburger2013realization,miyake2013realizing, PhysRevLett120-243201, PhysRevLett103-033003, martin2013quantum}. When neutral atoms are trapped in periodic potentials produced by standing wave of light fields, the trapping potentials of various atomic internal states are manipulated by lattice polarizations, which is called the spin-dependent optical lattice~\cite{mandel2003coherent,duan2003controlling}, which bring more complicated geometry for ultracold atoms such as spin-dependent hexagonal lattice~\cite{soltan2011multi}, spin-dependent optical superlattice~\cite{yang2017spin}, and have been used to study many interesting phenomena such as controlled coherent transport~\cite{ke2018compact,mandel2003coherent}, spinor BEC~\cite{ostrovskaya2004localization}, spin-orbit coupling and artificial gauge fields~\cite{ye2018band,grusdt2017tunable}, spontaneous emission of matter waves~\cite{krinner2018spontaneous}, twisted-bilayer optical potentials~\cite{PhysRevA1002019}.


Tune-out wavelengths that ac Stark shift cancels was initially introduced in species-specific optical manipulation~\cite{leblanc2007species} and can be useful for optical Feshbach resonances~\cite{PhysRevLett.115.155301} and atomic interferometer~\cite{PhysRevLett.114.140404}. Since tune-out wavelengths are independent of the light intensity \cite{arora2011tune,jiang2013tune-out,cheng2013tune}, it can be precisely measured by various methods~\cite{holmgren2012measurement,henson2015precision,leonard2015high,Schmidt2016,Adam2016Obtaining,trubko2017potassium,Copenhaver2019,Measurement2020}. In general, tune-out wavelength is utilized accurately only for the scalar shift by cancelling and neglecting the vector and tensor contribution as much as possible~\cite{holmgren2012measurement,henson2015precision,leonard2015high,trubko2017potassium}. In this paper, we investigate the tune-out wavelengths in more general cases of considering the contributions from both the scalar and vector shift. The ac Stark shift of optical lattice is measured by Kapitza-Dirac scattering which diffracts BEC into a number of high momentum states and the characteristics of spin-dependent optical lattice are investigated by scanning the lattice wavelength between the D1 and D2 lines. Kapitza-Dirac scattering becomes a standard tool and shows many applications in calibrating the lattice depth \cite{Ovchinnikov1999,cahn1997time,denschlag2002bose,beswick2019lattice,chen2010pulse}, detecting the lattice structure \cite{Viebahn2019,wen2020observation}, performing high-resolution spectroscopy \cite{Stenger1999} and metrology \cite{Gupta2002,Campbell2005}. The periodic potential originated from scalar shift and vector shift is manipulated by controlling the lattice polarizations, which is used to generate spin-dependent optical lattice. We can design the special spin-dependent optical lattice with the help of tune-out wavelengths.

\section{theory}
\subsection{AC Stark shift}
As we know, ac Stark effect is the result of an interaction between atoms and a classical light field. Here the total ac Stark shift for alkali-metal atoms in ground state interacting with a far-off-resonance light field can be expressed in terms of its scalar, vector, tensor components~\cite{beloytheory,lundblad2010experimental,le2013dynamical,becher2018anisotropic,tsyganok2019scalar}
\begin{equation}\label{eq:1}
	\begin{aligned}
		\Delta U=&\Delta U\left( F,{{m}_{F}};\omega  \right)\\
		=&-A[\alpha _{~}^{\left( 0 \right)}\left(\omega  \right)
		+{{\alpha }^{\left( 1 \right)}}\left(\omega  \right) \left( \xi \hat{e}_{k}\cdot\hat{e}_{B} \right) \frac{{m}_{F}}{F}\\
		& + {{\alpha }^{\left( 2 \right)}}\left(\omega  \right) \frac{3\cos^{2}\theta-1}{2} \frac{3{m}_{F}^{2}-F(F+1)}{F(2F-1)}],\\
	\end{aligned}
\end{equation}
where ${{\alpha }^{\left( 0,1,2 \right)}}\left(\omega  \right)$ are the scalar, vector and tensor polarizabilities respectively. $F$ is the total atomic angular momentum, $m_{F}$ is magnetic quantum number, $A$ is the laser field intensity with $A=2\epsilon_{0}c \left| E \right|^{2}$, $\omega$ and $E$ are the frequency and amplitude of optical field, $\hat{e}_{k}\cdot\hat{e}_{B}=\left\vert {e}_{k} \right\vert \left\vert {e}_{B} \right\vert \cos(\phi)$, $\hat{e}_{k}$ and $\hat{e}_{B}$ are unit vectors along the light wave-vector and magnetic field quantization axis respectively,  $\phi$ is the intersection angle between $\hat{e}_{k}$ and $\hat{e}_{B}$, $\theta$ is the intersection angle between the linearly polarized component of light field and $\hat{e}_{B}$. This formula comes from the first non-vanishing term (the second order) of a perturbation development. Note that the range of values of light ellipticity is $\xi\in[-1,1]$, $\xi=\pm 1$ denotes left and right circular polarization. The left and right elliptical polarization is defined in terms of the light wave-vector. Scalar shift can be interpreted as a spin-independent light shift. Vector shift acts like an effective magnetic field to generate the linear Zeeman splitting (light shift proportional to $m_{F}$), which depends on the ellipticity of the light and the intersection angle between the laser beam wave vector and magnetic field quantization axis $\hat{e}_{B}$. Tensor shift is proportional to $m^{2}_{F}$. For alkali-metal atoms in ground state, the tensor shift can vanish once light detuning $\delta$ exceed the hyperfine splitting $\Delta_{HF}$. Because the ground state is $J=1/2$, which induces the tensor shift coefficient ${{\alpha }^{\left( 2 \right)}}=0$~\cite{becher2018anisotropic,steck2007quantum,rosenbusch2009}. In this work, we consider a far detuning of $\delta\gg \Delta_{HF}$, so that the ac Stark shift only includes two terms of the scalar $\alpha^{(0)}$ and vector shift $\alpha^{(1)}$.


For a linear polarized light beam($\xi=0$), vector shift vanishes, but scalar shift keeps. For circular polarization light, the left and right circular polarization can change the sign of vector shift to be positive or negative. Therefore, the different ac Stark shift of two spin states can be cancelled by controlling the ellipticity, or tuning the angle between $\hat{e}_{k}$ and $\hat{e}_{B}$ (even changing the strength of the external bias magnetic field when considering high order contribution~\cite{lundblad2010experimental,andrei2010,yang2016}), which is an important technique for the atomic clock and qubit for quantum computation.

\subsection{Scalar and vector shift}
\begin{figure}[!htb]
	\centering
	\includegraphics[width=3in]{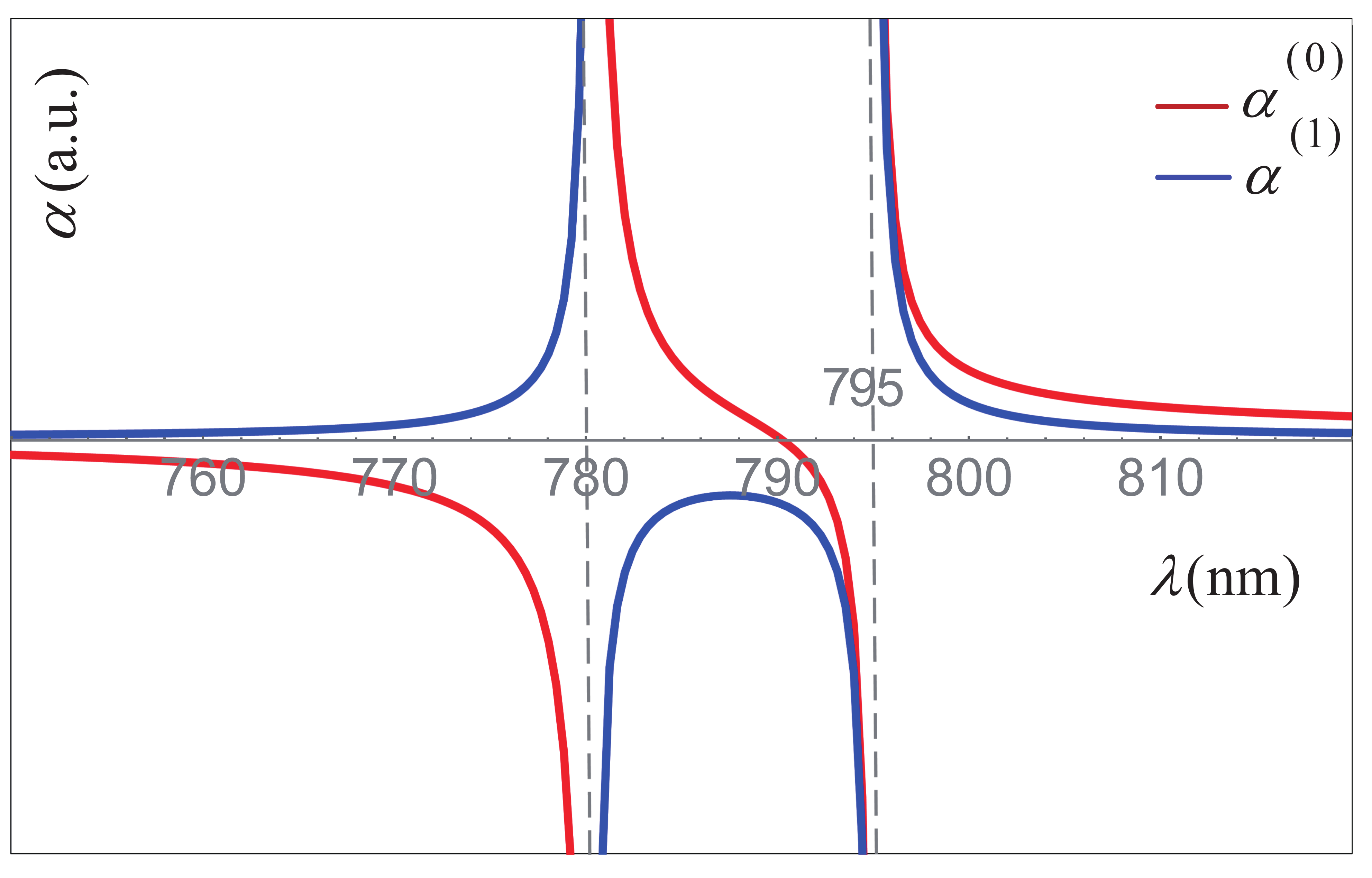}
	\caption{(Color online) \textbf{The coefficients of the scalar and vector shifts as the function of wavelength for $^{87}Rb$ atom in $|F=2,m_{F}=2\rangle$ state}.
		$\alpha^{(0)}$ has a crossed zero point at $\lambda$ = 790.020 nm. When the wavelength of light is far red-detuned from D1, $\alpha _{~}^{\left( 0 \right)}\gg \alpha _{~}^{\left( 1 \right)}\approx 0$. }
	\label{fig1}
\end{figure}

For the first excited state of alkali-metal atoms, the fine structure induces the spectral lines of the D1 (the $5^{2}S_{1/2}  \rightarrow5^{2}P_{1/2}$ transition) and D2 (the $5^{2}S_{1/2}  \rightarrow5^{2}P_{3/2}$ transition) lines. Because the D1 and D2 lines of the first excited state are larger detuned than the the excited-state hyperfine splitting, the coefficients of the scalar and vector shifts in Eq. (\ref{eq:1}) are expressed as ~\cite{corwin1999spin,park2001measurement,Mckay2009Thermometry,ming2012calculation}
\begin{equation}\label{eq:2}
	\begin{aligned}
		& { {{\alpha }^{\left( 0 \right)}}\left( \omega  \right)\approx -\frac{\pi {{c}^{2}}{{\Gamma }_{{{D2}}}}}{2\omega ^{3}_{0}}\left(\frac{2}{{{\delta }_{{{D2}}}}}+\frac{1}{{{\delta }_{{{D1}}}}} \right)}, \\
		& {{\alpha }^{\left( 1 \right)}}\left( \omega  \right)\approx -\frac{\pi {{c}^{2}}{{\Gamma }_{{{D2}}}}}{2\omega ^{3}_{0}}\left(\frac{1}{{{\delta }_{{{D2}}}}}-\frac{1}{{{\delta }_{{{D1}}}}} \right){g}_{F}{F},\\
		& {{\alpha }^{\left( 2 \right)}}\left( \omega  \right)\approx 0,\\
	\end{aligned}
\end{equation}
where ${{\Gamma }_{{{D}_{2}}}}$ is the decay rate of the excited state for $D_{2}$ line, ${{\delta }_{{{D}_{1}}}}={{\omega }_{~}}-{{\omega }_{D1}}$, ${{\delta }_{{{D}_{2}}}}={{\omega }_{~}}-{{\omega }_{D2}}$. ${{g}_{F}}$ is the gyromagnetic ratio
\begin{equation}\label{eq:2}
	\begin{aligned}
		&g_{F}=g_{J}\left[\frac{F(F+1)+J(J+1)-I(I+1)}{2F(F+1)}\right],\\
        &g_{J} =1+\frac{J(J+1)+S(S+1)-L(L+1)}{2J(J+1)},\\
	\end{aligned}
\end{equation}
where $S$ is the spin angular momentum, $L$ is the orbital angular momentum, $J$ is the total electronic angular momentum, $I$ is the total nuclear angular momentum. For the ground states $5^{2}S_{1/2}$ of $^{87}$Rb atoms, $g_{J}=2$, $g_{F}=1/2$ for $F=2$, $g_{F}=-1/2$ for $F=1$. Here, we study $^{87}Rb$ atoms and present the coefficients of the scalar and vector shifts as the function of wavelength in Fig.~\ref{fig1}. The resonant wavelengths of the D1 and D2 lines of $^{87}Rb$ atom are ${\lambda }_{D1}=794.98$ nm and ${\lambda }_{D2}=780.24$ nm respectively. Obviously $\alpha^{(0)}$ has a crossed zero point at $\lambda=790.005$ nm, and $\alpha^{(1)}$ always is negative between the D1 and D2 lines. When the wavelength of light is far red-detuned or blue-detuned by an amount larger than the fine structure splitting of the excited states, the vector shift approaches zero. Here, we study the tune-out wavelengths in more general cases of considering the contributions from both the scalar and vector shift. The tune-out wavelengths of the ground hyperfine states are given in Table \ref{tab:table1} with $\phi=0$, $\xi=0,\pm1$.

\begin{table*}[!htb]
    \renewcommand\arraystretch{1}
		\centering
	
    	\caption{Tune-out wavelengths of the ground states with $\phi=0$ in the $5^{2}S_{1/2}-5^{2}P_{1/2,}$ $_{3/2}$ states of $^{87}$Rb.}

        \begin{threeparttable}
        	
    	\begin{tabular}{cccccccccc}

    \hline
    \hline
		
				\toprule
				
				\multirow{2}{*}{Polarization} & \multirow{2}{*}{$|F,m_{F}\rangle$} & \multicolumn{4}{c}{${{\lambda_{zero}}}$(nm)} \\
				
				&                         &  Calc. use Eq. (\ref{eq:2})                     & Calc. in Ref~\cite{leblanc2007species}            &  Other Calc.      &   Expt.  \\

\hline
				
				\midrule
				
				&   $|2,2\rangle$         &   790.005       &  790.04                                 &                                                   & 790.01850(9)~\cite{Schmidt2016}                                \\
				&   $|2,    1\rangle$     &   790.005       &  790.04                                 &  790.032439(35)~\cite{leonard2015high}            &                                 \\
				$\xi=0$&   $|2,0\rangle$         &   790.005       &  790.03                                 &  790.034(7)~\cite{arora2011tune}           & 790.032388(32)~\cite{leonard2015high} \\
				&   $|2,-1\rangle$        &   790.005       &  790.04                                 &  790.032602(193)~\cite{Wang2016}         &                  \\
				&   $|2,-2\rangle$        &   790.005       &  790.04                                 &                                                   &                                \\
				&   $|1,1\rangle$         &   790.005       &  790.04                                 &                                                   &   789.85(1) ~\cite{Catani2009}    \\	
				&   $|1,0\rangle$         &   790.005       &  790.04                                 &  790.018187(193)~\cite{Wang2016}                  &   790.018(2)~\cite{Lamporesi2010}, 790.020(25) \tnote{*}            \\
				&   $|1,-1\rangle$        &   790.005       &  790.04                                 &                                                   &   790.01858(23)~\cite{Schmidt2016} \\
				\\
				&   $|2,2\rangle$         &   none          &  none                                   &                                                   &   none \tnote{*}                                 \\
				&   $|2,1\rangle$         &   792.484       &  792.52                                 &                                                   &                                  \\
				&   $|2,0\rangle$         &   790.005       &  790.06                                 &                                                   &                           \\
				$\xi=1$ &   $|2,-1\rangle$        &   787.541       &  787.59                                 &                                                   &                                  \\
				&   $|2,-2\rangle$        &   785.093       &  785.14                                 &                                                   &                                \\
				&   $|1,1\rangle$         &   787.541       &  787.59                                 &                                                   &   787.590(31) \tnote{*}                      \\
				&   $|1,0\rangle$         &   790.005       &  790.06                                 &                                                   &   790.020(25)  \tnote{*}             \\
				&   $|1,-1\rangle$        &   792.484       &  792.53                                 &                                                   &                                     \\
				\\
				&   $|2,2\rangle$         &   785.093       &  785.14                                 &  785.11516~\cite{Wang2017}                        &   785.146(12) \tnote{*}                            \\
				&   $|2,1\rangle$         &   787.541       &  787.59                                 &                                                   &                              \\
				&   $|2,0\rangle$         &   790.005       &  790.06                                 &                                                   &                        \\
				$\xi=-1$&   $|2,-1\rangle$        &   792.484       &  792.52                                 &                                                   &                                   \\
				&   $|2,-2\rangle$        &    none         &  none                                   &  none~\cite{Wang2017}                             &                                 \\
				&   $|1,1\rangle$         &   792.484       &  792.53                                 &                                                   &  $\approx$792.4~\cite{Schmidt2016}, 792.462(22) \tnote{*}              \\
				&   $|1,0\rangle$         &   790.005       &  790.06                                 &                                                   &    790.020(25) \tnote{*}                \\
				&   $|1,-1\rangle$        &   787.541       &  787.59                                 &                                                   &  $\approx$787.620~\cite{Schmidt2016}          \\
		\bottomrule	

\hline
	
		\end{tabular}

        \begin{tablenotes}
    	     \footnotesize
	         \item[*] Our experimental measurements.
        \end{tablenotes}
        \end{threeparttable}
    \label{tab:table1}
\end{table*}

In order to measure the tune-out wavelengths, we employ a one-dimensional (1D) optical lattice along the external bias magnetic field ($\phi=0$) with different polarization configurations. Here, the two laser beams have the same intensity with $A=2\epsilon_{0}c\left| E_{1} \right|^{2}=2\epsilon_{0}c\left| E_{2} \right|^{2}$. For case 1, two laser beams with the same linear polarization counter-propagate along the z axis. Because of the parallel polarized beams, it can produce the spatial intensity modulation to form a 1D optical lattice,
\begin{equation}\label{eq:4}
	\begin{split}
		\Delta U_{L1}\left( F,{{m}_{F}};\omega  \right)=&-4A  {\alpha }^{\left( 0 \right)}\cos^{2}kz.\\
	\end{split}
\end{equation}
The optical lattice potential for this case only originates from the scalar shift. Therefor it can be used for measuring the tune-out wavelength for the scalar shift. For case 2, two counter-propagated laser beams have linear orthogonal polarization (lin$\perp$lin polarization configuration). The orthogonally polarized beams can not produce a spatial intensity modulation. In contrast, it can produce the ellipticity modulation of polarization in space. This optical lattice potential is called Sisyphus optical potential, which has been used for Sisyphus cooling~\cite{dalibard1989laser}. The periodic potential is given by
\begin{equation}\label{eq:5}
	\begin{split}
		\Delta U_{L2}\left( F,{{m}_{F}};\omega  \right)=&-2A [ {\alpha }^{\left( 0 \right)}+{\alpha }^{\left( 1 \right)}\xi\frac{{m}_{F}}{F}\sin (2kz)].\\
	\end{split}
\end{equation}
This periodic potential only comes from the vector shift and the scalar term, which gives a uniformed energy shift. For case 3, two laser beams with the same polarized circular polarization counter-propagates along the z axis, which can also produce the spatial intensity modulation to form a 1D optical lattice,
\begin{equation}\label{eq:6}
	\begin{split}
		\Delta U_{L3}\left( F,{{m}_{F}};\omega  \right)=&-4A\cos^{2} kz [ {\alpha }^{\left( 0 \right)}+{\alpha }^{\left( 1 \right)}\xi\frac{{m}_{F}}{F}].\\
	\end{split}
\end{equation}
The optical lattice potential for this case includes the scalar and vector shift simultaneously. Therefore, we can study the tune-out wavelengths in presence of contributions of both the scalar and vector shift in this case. And for case 4, like case 2, the scalar shift is a constant only related to the constant intensity of two orthogonal polarized counter-propagating laser beams. But the different with former, is that the vector shift becomes zero since it only produce the rotation of linear polarization in space. Hence it can not produce any spatial modulation to form a 1D lattice.
\begin{equation}\label{eq:7}
	\begin{split}
		\Delta U_{L4}\left( F,{{m}_{F}};\omega  \right)=&-2A {\alpha }^{\left( 0 \right)}.\\
	\end{split}
\end{equation}
Here, for a 1D optical lattice along the external bias magnetic field ($|\cos\phi|=1$), it is convenient to define the left and right elliptical polarizations in terms of the magnetic field quantization axis.

\begin{figure*}[!htb]
	\includegraphics[width=6.8in]{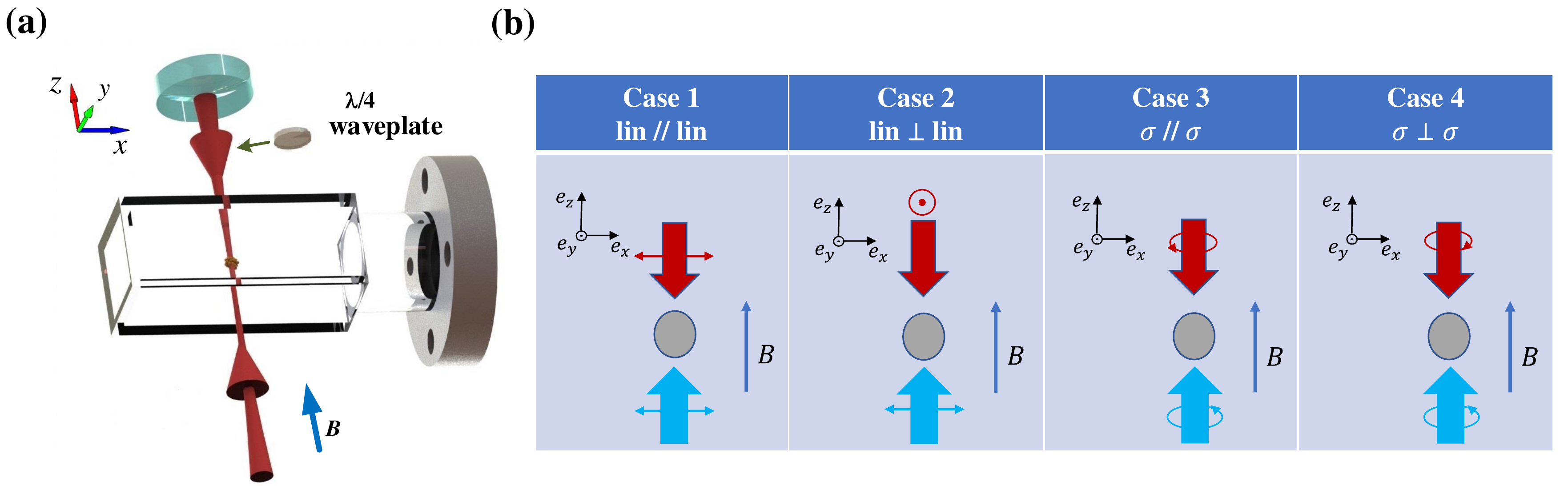}
	\centering
	\caption{(Color online) \textbf{Schematic of experimental setup and three cases for different polarization configurations}.
		\textbf{(a)}A 1D optical lattice is formed by two counter-propagating laser fields. The external magnetic field is aligned along with -z axis.
		\textbf{(b)} Case 1, the counter-propagating lasers have  linear parallel polarization. Case 2, the counter-propagating lasers have linear orthogonal polarization. Case 3 and Case 4, the counter-propagating lasers have the circular parallel or orthogonal polarizations.}
	\label{fig2}
\end{figure*}

\section{Experiment}
The schematic of experimental setup is shown in Fig.~\ref{fig2}(a). Ultracold $^{87}$Rb atoms in $|F=2,{{m}_{F}}=2\rangle$ hyperfine state are loaded in a crossed optical dipole trap \cite{Xiong:10}. Forced evaporation in the optical trap is used to create the BEC with up to $5\times10^5$ atoms. In order to obtain the atoms in the different single spin states, BEC is transferred from $|F=2,{{m}_{F}}=2\rangle$ into $|F=1,{{m}_{F}}=1\rangle$ via a rapid adiabatic passage induced by a microwave-frequency field with duration of 10 ms at 3.9 G of the bias magnetic field, where the frequency of center is 6.842935 GHz and the width is 0.25 MHz. Then the atoms in $|F=1,{{m}_{F}}=1\rangle$ can further be transferred into $|F=1,{{m}_{F}}=0\rangle$ state using a rapid adiabatic passage induced by a radio frequency (rf) field at 28 G of the bias magnetic field.
The lattice beam is derived from a single frequency Ti:sapphire laser with broad tuning range of the frequency. An acousto-optical modulator is used to control the intensity of the lattice beam. The lattice beam passes through the polarizing beam splitter to generate the perfect polarization. The polarization extinction ratio of the polarizing beam splitter can reach 500:1. Therefore the linear polarization purity of the lattice beam is about 0.2$\%$. Furthermore the circular polarization purity of the lattice beam can reach about 0.5$\%$. A lattice beam propagates with z axis and converges on BEC with waist of 100 $\mu m$ by a lens (f = 300 mm). Then the beam is reflected by a concave mirror ( curvature radius r=300 mm) and refocused on BEC with the almost same waist size. The advantage of this configuration can reduce phase jitter significantly. Here, we employ Kapitza-Dirac (or Raman-Nath) scattering to measure the ac Stark shift. Kapitza-Dirac scattering is used to diffract the BEC into a number of momentum states by a standing light wave, in which the interaction is sufficiently short and strong \cite{gould1986diffraction}. In this process, BEC is kept in a crossed optical dipole trap and the lattice potential imprints a phase modulation on matter wave in position space. Then the phase modulation on matter wave is measured in momentum space via the time-of-flight (TOF) absorption image. It is obvious that higher momentum orders $\pm2N\hbar k$ appear in the atomic density distribution of the TOF absorption image, which depends on the potential depth and interaction time. Here, we apply a 1D optical lattice short pulse for 4 $\mu s$ with the power of 80 mW on BEC. Then immediately turn off the optical trap, let the atoms ballistically expand in 12 ms and take the absorption images. We obtain the lattice depth from the absorption images by applying the lattice at a fixed laser power for different intervals of time and by observing the interval at which the $n=0$ order atoms in the lattice vanish \cite{Ovchinnikov1999,cahn1997time,denschlag2002bose,beswick2019lattice,chen2010pulse}. We define the recoil momentum $\hbar k=2\pi
\hbar /\lambda$ and recoil energy $E_{r}=(\hbar k)^{2}/2m =  h\times3.67$ kHz as the nature momentum and energy
units, where $m$ is the mass of $^{87}$Rb atom, $\lambda$ is the wavelength of the lattice laser.

\begin{figure*}[!htb]
	\centering
	\includegraphics[width=6.9in]{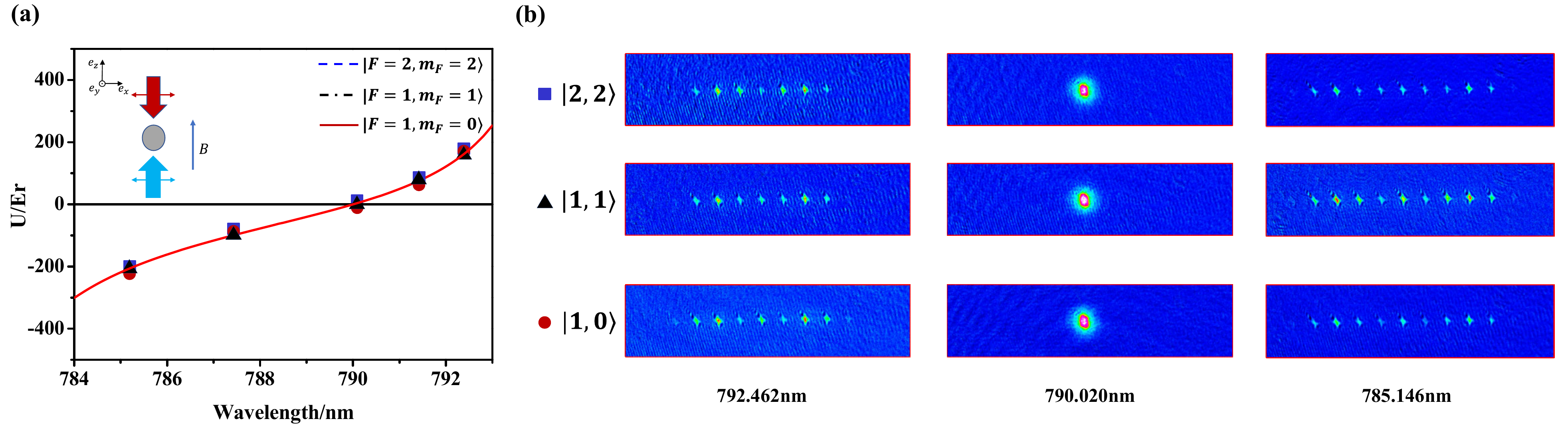}
	\caption{(Color online)
		\textbf{The 1D optical lattice with the linear parallel polarization}. (a) Measured data (squares, triangles and circles) and the theoretical fit (the three types of lines overlapped together) of the lattice potential depth as the function of the laser wavelength for the three hyperfine states $|F=2,{{m}_{F}}=2\rangle$, $|F=1,{{m}_{F}}=1\rangle$ and $|F=1,{{m}_{F}}=0\rangle$. It shows that this periodic potential is a spin-independent lattice. Each
point is the average of at least three measurements. (b) Atomic density distribution in the TOF absorption images at different wavelength of the lattice laser.}
	\label{fig3}
\end{figure*}

\begin{figure*}[!htb]
	\centering
	\includegraphics[width=6.9in]{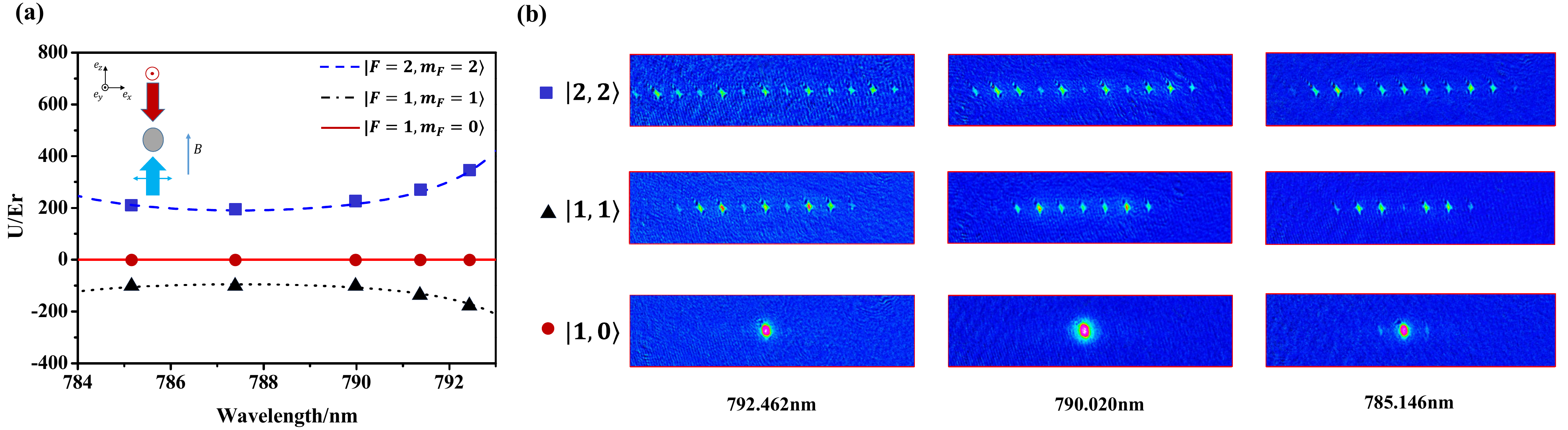}
	\caption{(Color online)
		\textbf{The 1D optical lattice with the linear orthogonal polarization}. (a) Measured data (squares, triangles and circles) and the theoretical fit (dashed, solid and dotted lines) of the lattice potential depth as the function of the laser wavelength for the three hyperfine states $|F=2,{{m}_{F}}=2\rangle$, $|F=1,{{m}_{F}}=1\rangle$ and $|F=1,{{m}_{F}}=0\rangle$. This periodic potential is a spin-dependent lattice, which only depends on the vector shift. (b) Atomic density distribution in the TOF absorption images at different wavelength of the lattice laser. There is no lattice potential for the $|F=1,{{m}_{F}}=0\rangle$ state ($
		\Delta{{m}_{F}}=0$) while the $|{{m}_{F}}=\pm 2\rangle$ and $|{{m}_{F}}=\pm 1\rangle$ states ($
		\Delta{{m}_{F}}\neq0$) always experience the lattice potential when adjusting the lattice wavelength.}
	\label{fig4}
\end{figure*}

\begin{figure*}[!htb]
	\centering
	\includegraphics[width=6.9in]{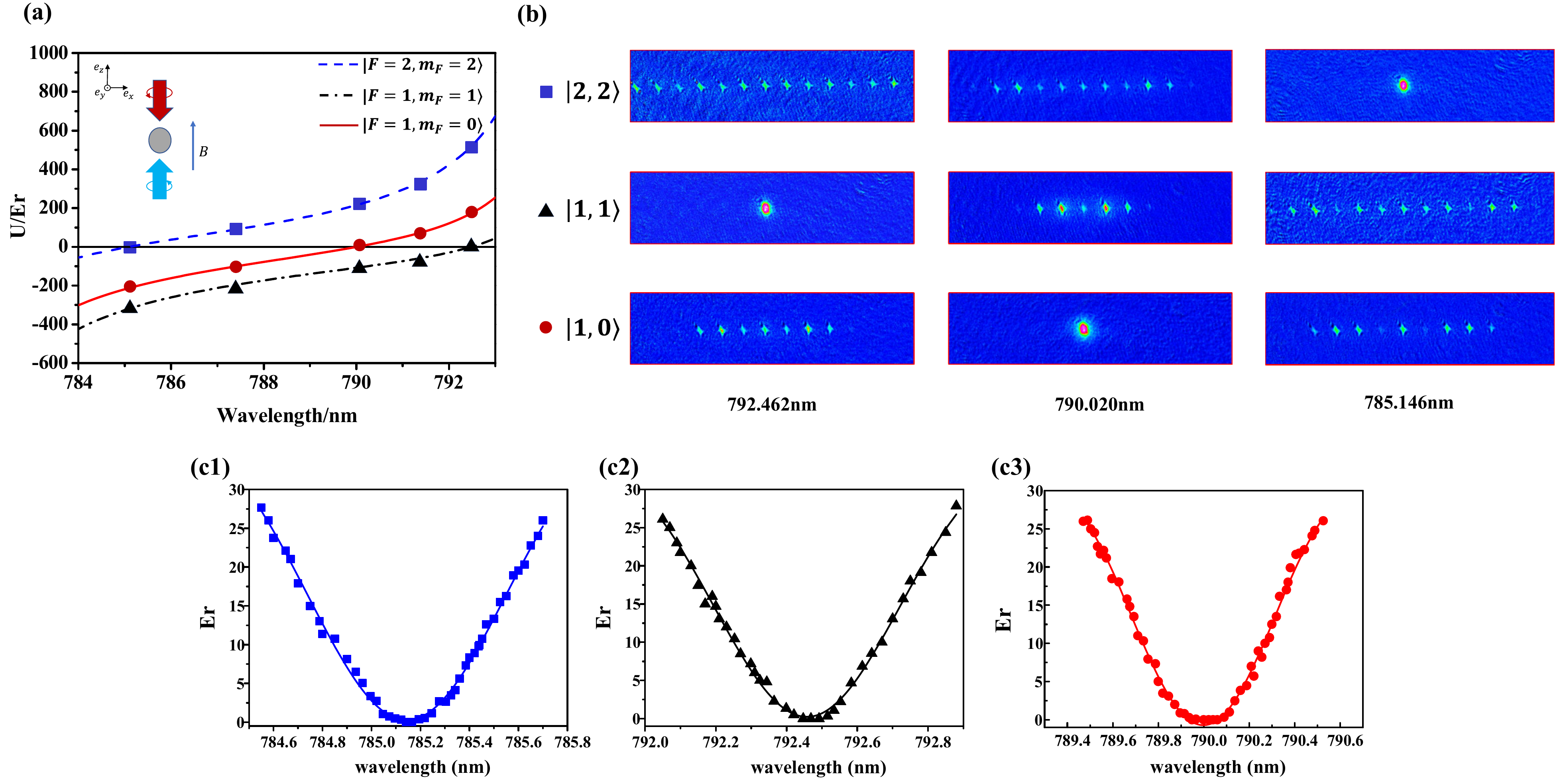}
	\caption{(Color online)
		\textbf{The 1D optical lattice with the circular parallel polarization}. (a) Measured data (squares, triangles and circles) and the theoretical fit  (dashed, solid and dotted lines) of the lattice potential depth as the function of the laser wavelength for the three hyperfine states $|F=2,{{m}_{F}}=2\rangle$, $|F=1,{{m}_{F}}=1\rangle$ and $|F=1,{{m}_{F}}=0\rangle$. This periodic potential is a spin-dependent lattice, which depends on the the scalar and vector shift. The tune-out wavelengths are 792.462(22) nm, 790.020(25) nm and 785.146(12) nm for the $|F=1,{{m}_{F}}=1\rangle$, $|F=1,{{m}_{F}}=0\rangle$ and $|F=2,{{m}_{F}}=2\rangle$ respectively. (b)  Atomic density distribution in the TOF absorption images at different wavelength of the lattice laser. (c1)-(c3) give the detailed measurements of the lattice potentials for the $|F=1,{{m}_{F}}=1\rangle$, $|F=1,{{m}_{F}}=0\rangle$ and $|F=2,{{m}_{F}}=2\rangle$ near the tune-out wavelengths respectively.}
	\label{fig5}
\end{figure*}

\begin{figure*}[!htb]
	\centering
	\includegraphics[width=6.9in]{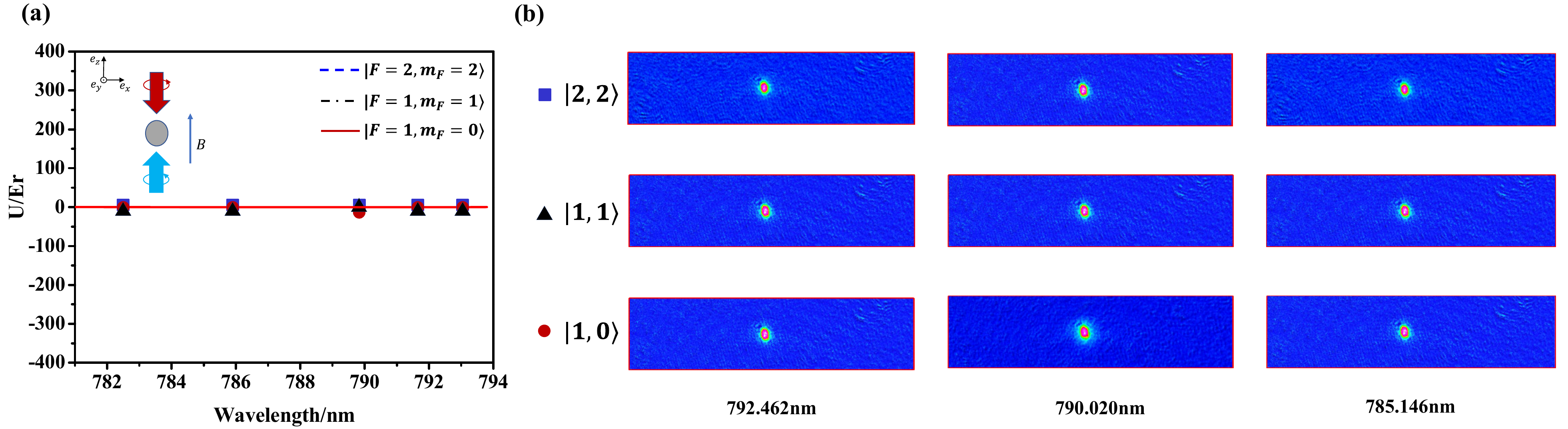}
	\caption{(Color online) \textbf{The 1D optical lattice with the circular orthogonal polarization}. (a) Measured data (squares, triangles and circles) and the theoretical fit (the three types of lines overlapped together) of the lattice potential depth as the function of the laser wavelength for the three hyperfine states $|F=2,{{m}_{F}}=2\rangle$, $|F=1,{{m}_{F}}=1\rangle$ and $|F=1,{{m}_{F}}=0\rangle$. There is no obvious periodic potential in this spin-dependent lattice, because the scalar and vector don't produce spatial light intensity modulation. (b) Atomic density distribution of the time-of-flight absorption images. There is no lattice potentials for all states for all wavelength.}
	\label{fig6}
\end{figure*}

For case 1, a 1D optical lattice with the linear parallel polarization produces the spatial intensity modulation, which only comes from the scalar shift. Thus it is a spin-independent optical lattice and the potential depth as the function of the lattice wavelength are plotted in Fig.~\ref{fig3}(a). Here, the positive and negative periodic potentials correspond to blue and red detuned lattice laser. The higher momentum orders $\pm2N\hbar k$ are observed in the atomic density distribution of the TOF absorption images as shown in Fig.~\ref{fig3}(b), which depends on the potential depth. We measure the tune-out wavelengths by changing the lattice wavelength and find its location at $\sim$790.020 nm for all spin states, which is in good agreement with the previous works ~\cite{leonard2015high,Schmidt2016}.

\begin{figure*}[!htp]
	\centering
	\includegraphics[width=7in]{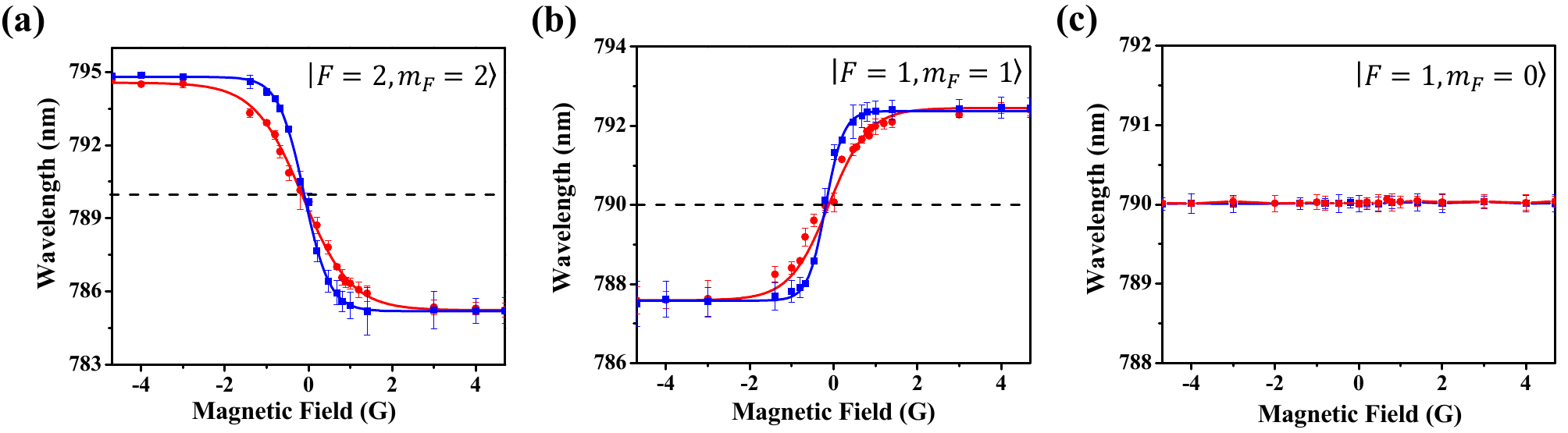}
	\caption{(Color online) \textbf{The tune-out wavelengths as the function of the amplitude of the external magnetic field for Case 3.}
		(a), (b), (c) correspond to  $|F=2,{{m}_{F}}=2\rangle$, $|F=1,{{m}_{F}}=1\rangle$ and $|F=1,{{m}_{F}}=0\rangle$. Blue (square sign) and red (circular sign) curves correspond to the residual magnetic field in the (x, y, z) directions suppressed from (0.4 G, 0.46 G, 0.38 G) to (0.25 G, 0.28 G, 0.2 G) respectively. }
	\label{fig7}
\end{figure*}

For case 2, a laser beam with the linear polarization pass through a quarter-wave plate and is reflected by a concave mirror, which produces the linear orthogonal polarization configuration. We plot the potential depth vs different wavelength as shown in Fig.~\ref{fig4}(a). This periodic potential is a spin-dependent lattice, which only comes from the contribution of the vector shift. Therefore, there is no lattice potential for the $|F=1,{{m}_{F}}=0\rangle$ state ($
\Delta{{m}_{F}}=0$) for all wavelength as shown in Fig.~\ref{fig4}(b). The $|{{m}_{F}}=\pm 2\rangle$ and $|{{m}_{F}}=\pm 1\rangle$ states ($
\Delta{{m}_{F}}\neq0$) always experience the lattice potential when adjusting the lattice wavelength.

For case 3, a 1D optical lattice with the circular parallel polarization produces the spatial intensity modulation, which includes the contribution of the scalar and vector shift simultaneously. The potential depth of $|F=1,{{m}_{F}}=1\rangle$, $|F=1,{{m}_{F}}=0\rangle$ and $|F=2,{{m}_{F}}=2\rangle$ are given in Fig.~\ref{fig5}(a) and (b). The tune-out wavelengths are generated by balancing the scalar and vector shift, which are 792.462(22)nm, 790.020(25)nm and 785.146(12)nm for $|F=1,{{m}_{F}}=1\rangle$, $|F=1,{{m}_{F}}=0\rangle$ and $|F=2,{{m}_{F}}=2\rangle$ respectively as shown in Fig.~\ref{fig5}(c). Note that the tune-out wavelengths in this case are sensitive to the ellipticity of the lattice polarization, the alignment between the direction of the 1D optical lattice and the external bias magnetic field. This case provides us more controlled ways to generate the different kinds of spin-dependent optical lattice. There is an interesting phenomenon that two neighbouring spin states have the opposite lattice potential (blue and red detuning respectively) by choosing the appropriate wavelength, for example, $|F=1,{{m}_{F}}=1\rangle$ and $|F=1,{{m}_{F}}=0\rangle$ states have the opposite lattice potential at the wavelength of 791.24 nm.

For case 4, two counter-propagating laser beams with the orthogonal circular polarization can not generate any spatial modulation on the BEC. The potential depth for $|F=1,{{m}_{F}}=1\rangle$, $|F=1,{{m}_{F}}=0\rangle$ and $|F=2,{{m}_{F}}=2\rangle$ are given in Fig.~\ref{fig6}(a) and (b), showing no effective potential (no density modulation) for these states.

Furthermore, we study the dependence of the tune-out wavelengths on the strength of the external bias magnetic field in more detail. The intersection angle $\phi$ between $\hat{e}_{k}$ and $\hat{e}_{B}$ is expressed as
\begin{equation}\label{eq:8}
	\begin{split}
		cos(\phi)=\frac{B_{Bi}^{z}+B_{Re}^{z}}{\sqrt{(B_{Re}^{x})^{2}+(B_{Re}^{y})^{2}+(B_{Bi}^{z}+B_{Re}^{z})^{2}}},\\
	\end{split}
\end{equation}
where $B_{Bi}$ is the external bias magnetic field, $B_{Re}$ is the residual magnetic field (such as the earth magnetic field). We measure the tune-out wavelengths as the function of the strength of the external bias magnetic field as shown in Fig.~\ref{fig7}. By changing the bias magnetic field to small value, the direction of the total magnetic field and the intersection angle $\phi$ can be changed. Therefore, the tune-out wavelengths change when the strength of the bias magnetic field is near to the residual magnetic field value. The strength of the external bias magnetic field in one direction is gradually decreased to zero and then increased in the opposite direction. We find that the tune-out wavelengths for $\xi=1$ jumps into $\xi=-1$ due to inversion the external bias magnetic field direction for the spin $|F=1,{{m}_{F}}=1\rangle$ and $|F=2,{{m}_{F}}=2\rangle$ states as shown in Fig.~\ref{fig7}(a) and (b). The slope is sensitive to the strength of residual magnetic field in the perpendicular direction of z axis. Here, three pairs of Helmholtz coils are employed to compensate the background magnetic field. When the residual magnetic field in the (x, y, z) directions are suppressed from (0.4 G, 0.46 G, 0.38 G) to (0.25 G, 0.28 G, 0.2 G) which are measured by a triaxial fluxgate magnetometer, the slope is changed from -4.054 nm/G to -6.584 nm/G for $|F=1,{{m}_{F}}=1\rangle$ and 8.108 nm/G to 13.168 nm/G for $|F=2,{{m}_{F}}=2\rangle$ as shown in Fig.~\ref{fig7}(a) and (b) respectively. Therefore, this method can be utilized to calibrate and measure the residual magnetic field.

\section{Conclusion}
In conclusion, we present an experiment to measure the ac Stark shift around the tune-out wavelengths of $^{87}$Rb BEC in the three different hyperfine ground states $|F=1,{{m}_{F}}=1\rangle$, $|F=1,{{m}_{F}}=0\rangle$ and $|F=2,{{m}_{F}}=2\rangle$ between D1 and D2 lines. Four different polarization configurations of one-dimensional optical lattice, which are originated from scalar shift, vector shift, both scalar and vector shift respectively by manipulating the lattice polarizations, are investigated. Kapitza-Dirac scattering technique is employed to probe the ac Stark shift of atoms in optical lattice and the characteristics of spin-dependent optical lattice are presented by scanning the lattice wavelength. We present the tune-out wavelengths in more general cases of considering the contributions of both the scalar and vector shift. We further study the dependence of the tune-out wavelengths on the strength of the external bias magnetic field in more detail. Our work provides a clear interpretation for spin-dependent optical lattice and can be used for the realization of two species system, or the same atoms (Rb) with different spin states, in which one of them move freely while the others are trapped to different degrees of optical lattice potential. This system can be a test-bed for observing or simulating phenomena such as entropy cooling \cite{,Catani2009,Lamporesi2010,Mckay2009Thermometry,arora2011tune,Bause2020}, Kondo effect \cite{ren2016, yao2019} etc, and even have application in Sr optical lattice clock \cite{Heinz2020}.

\section{Funding Information}
This research was supported by the MOST (2016YFA0301602, 2018YFA0307601), NSFC (Grant No. 11804203, 11974224, 11704234), the Fund for Shanxi $\text{"1331 Project"}$ Key Subjects Construction.




\bibliography{references}

\end{document}